\documentclass[a4paper,11pt]{article}
\usepackage{pos}

\usepackage{graphicx}

\usepackage{pstricks,pst-node,pst-text,pst-3d}
\usepackage{pst-all}

\usepackage{subcaption}
\usepackage{svg}
\usepackage{amsmath}
\usepackage{booktabs}
\usepackage{multirow}
\usepackage{tikz}

\def\lsim{\raise0.3ex\hbox{$\;<$\kern-0.75em\raise-1.1ex\hbox{$\sim\;$}}}
\def\gsim{\raise0.3ex\hbox{$\;>$\kern-0.75em\raise-1.1ex\hbox{$\sim\;$}}}
\def\eps{\varepsilon}
\def\theta{\vartheta}

\def\Xmax{$X_{\max}$ }
\def\X2{$\sigma(X_{\max})$}

\def\d{{\rm d}}
\def\be{\begin{equation}}
\def\ee{\end{equation}}
\def\ba{\begin{align}}
\def\ea{\end{align}}

\newcommand{\EeV}{\mathrm{EeV}}

\newcommand{\nat}{{Nature }}

\newlength{\bibitemsep}
\newlength{\bibparskip}
\let\oldthebibliography\thebibliography
\renewcommand\thebibliography[1]{%
  \oldthebibliography{#1}%
  \setlength{\parskip}{\bibitemsep}%
  \setlength{\itemsep}{\bibparskip}%
}

\title{Extragalactic Cosmic Rays}

\ShortTitle{Extragalactic Cosmic Rays}

\author{M.~Kachelrie\ss}

\affiliation{Institutt for fysikk, NTNU, Trondheim, Norway}

\abstract{
  I review the status of ultrahigh-energy cosmic ray (UHECR) physics.
  After introducing the main
  experimental results and summarizing possible intepretations, I discuss
  observational and theoretical constraints on the sources of UHECRs. I
  comment also briefly on the role of magnetic fields. Combining these
  constraints, I argue that luminuous and numerous AGN types as FR-I and
  Seyfert galaxies, or alternatively hypernovae, are the most promising
  UHECR sources. Finally,
  I sketch few of the models presented at the conference before concluding.
}

\FullConference{37th International Cosmic Ray Conference -ICRC2021-\\
                15-22 July 2021\\
                Online -- Berlin, Germany}

\begin{document}


\maketitle

\section{Introduction}

A review with the title ``Extragalactic Cosmic Rays'' requires first to
clarify the lowest energies to be included, what in turn depends
on the question where the transition between Galactic and extragalactic
CRs takes place. I will later argue that this transition happens
around $5\times 10^{17}$\,eV. To be definite, I call ultra-high
energy cosmic rays (UHECR) all particles with energies above  $10^{17}$\,eV. 
I start with a discussion of the basic experimental results in Sec.~2,
before I comment briefly on the role of magnetic fields in Sec.~3.
Then observational and theoretical constraints on the sources of UHECRs
are discussed in Sec.~4.  Finally, I sketch a small selection of the UHECR
models presented at this ICRC.

\section{Observations and their interpretation}

\subsection{Energy spectrum}

\begin{figure}[b]
  \centering
  \includegraphics[width=0.49\columnwidth]{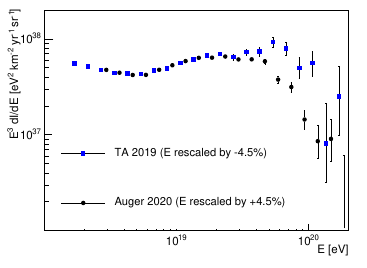}
  \includegraphics[width=0.49\columnwidth]{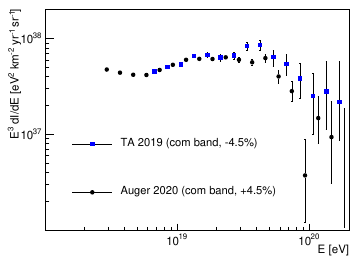}
  \caption{UHECR spectrum measured by the PAO (Auger) and TA experiments
    after a constant relative shift by 9.0\%; 
  Left panel full field of view and right panel
  in the common declination band;
  from Ref.~\protect\cite{Tsunesada:2021qO}. }
\label{fig:UHECR_spectrum}
\end{figure}

In Fig.~\ref{fig:UHECR_spectrum}, measurements of the UHECR intensity
$I(E)$ from the two experiments with the largest exposure, the Pierre Auger
Observatory (PAO) and the  Telescope Array (TA) are
presented~\cite{Tsunesada:2021qO}. The uncertainty in the absolute energy
scale of
these experiments is around 10\%, leading to large shifts in a plot of
$E^3I(E)$. The two experiments can be cross-calibrated using the fact that the
UHECR intensity up to $10^{19}$\,eV is very isotropic: Applying a relative
shift of the energy scale of the two experiments, their all-particle
intensities shown in the left panel of Fig.~\ref{fig:UHECR_spectrum} averaged
over their full field of view agree well up to $3\times 10^{19}$\,eV.
At higher energies, the deviations increase, with the cutoff in the TA
spectrum shifted to higher energies.  The agreement at the highest energies
improves in the declination band common to the two experiments shown
in the right panel, but some differences persists. Their
elimination would require an energy-dependent
rescaling. In addition, TA sees  slightly
different spectra in the southern and northern part of its sky. 
Apart from unaccounted systematic effects, the imprint of the large-scale
structure (LSS) on the source distribution which is not averaged out at
these energies could explain this difference. Moreover,
variations in the maximal energies of the most important sources in
different parts of the sky can become important. Alternatively,
when the cutoff is caused by the GZK effect, differences in the distance
to the dominating sources may cause an intensity difference.
Note that the energy $E_{1/2}\simeq 5\times 10^{19}$\,eV where the integral
intensity drops for a pure proton composition by a factor two relative to
the intensity
expected without pion production~\cite{Berezinsky:2002nc} deviates
significantly from the value determined from the PAO data,
$E_{1/2}\simeq (2.3\pm 0.4) \times 10^{19}$\,eV. Such a low value of $E_{1/2}$
points to an intermediate mass composition at the highest energies,
if the suppression is a propagation effect.

\begin{figure}
\hspace*{-0.5cm}  
\includegraphics[width=0.52\columnwidth,angle=0]{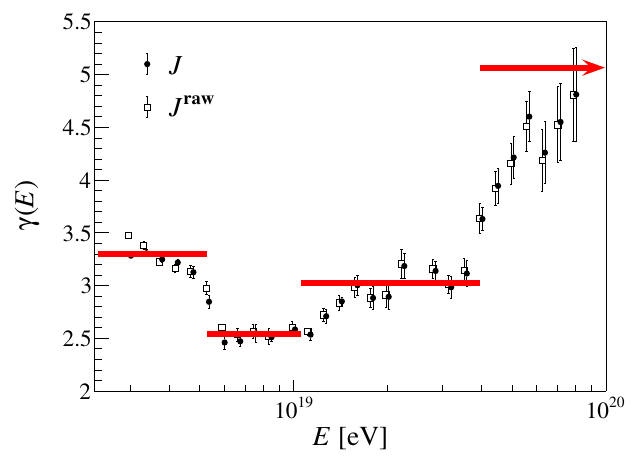}
\hfill
\includegraphics[width=0.5\columnwidth,angle=0]{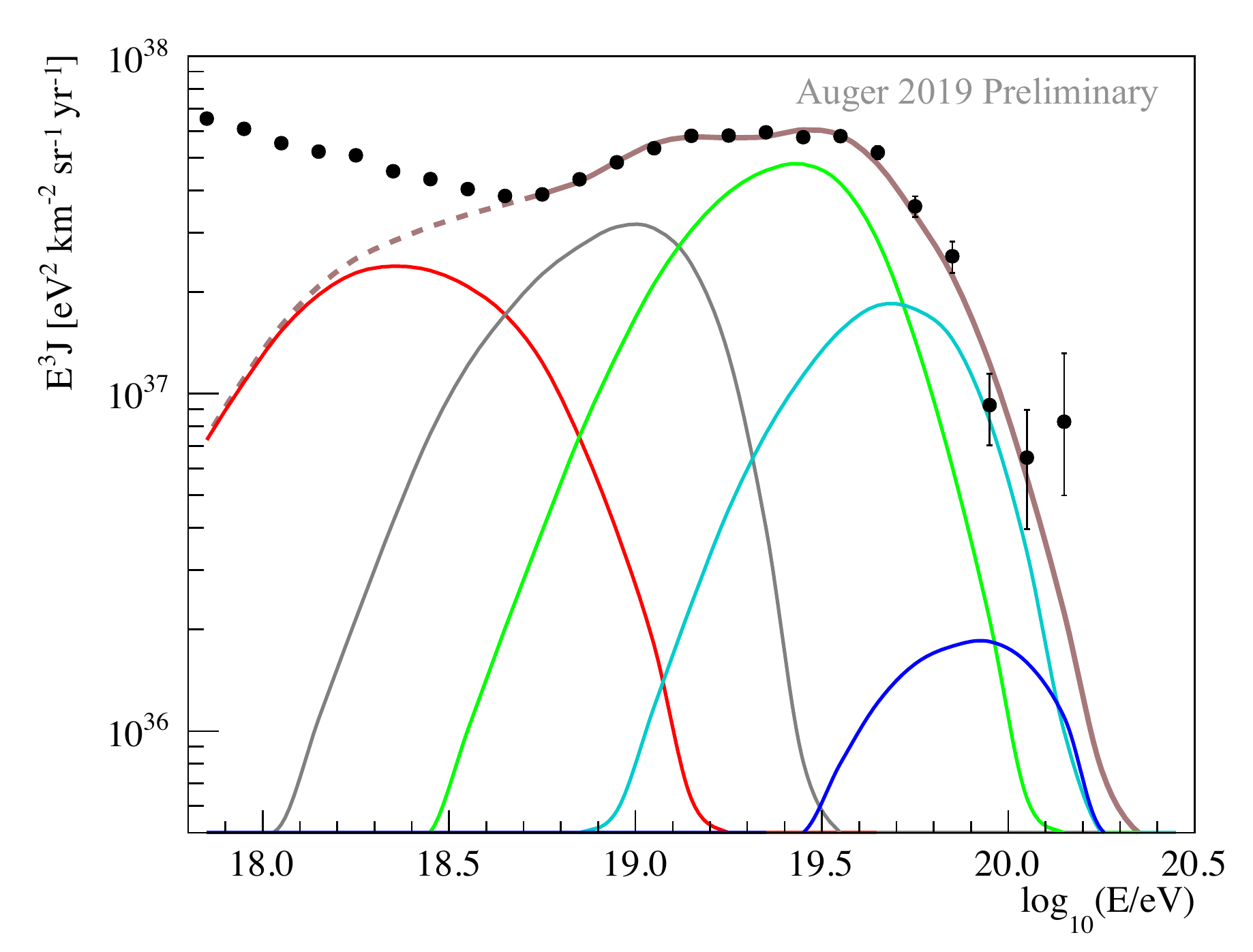}
\caption{Left: Local fit of power-law index to the CR intensity measured by the
  PAO compared to the index obtained using ``four segments'' (red line). Right:
  A model fit to both spectrum and composition data~\cite{Castellina:2019huz}.
\label{fitspec}}
\end{figure}

The energy spectrum of CRs above  $10^{17}$\,eV shows several features.
The second knee around $E\simeq 5\times10^{17}$\,eV,
the ankle at $E\simeq3\times 10^{18}$\,eV and a flux suppression at the
highest energies. An additional feature, the so-called instep, has been
recently established by the
PAO~\cite{PierreAuger:2020kuy,PierreAuger:2020qqz}.
In the left panel of Fig.~\ref{fitspec}, the slopes obtained fitting
a power-law with four segments to the data are compared to
the varying slope derived by fitting the spectrum locally including
three to six energy bins. This comparison
shows clearly that, given the small experimental errors, broken power laws
are not an adequate description of the energy spectrum.

\subsection{Composition}

\begin{figure}[tb]
  \centering
  \includegraphics[width=0.52\linewidth]{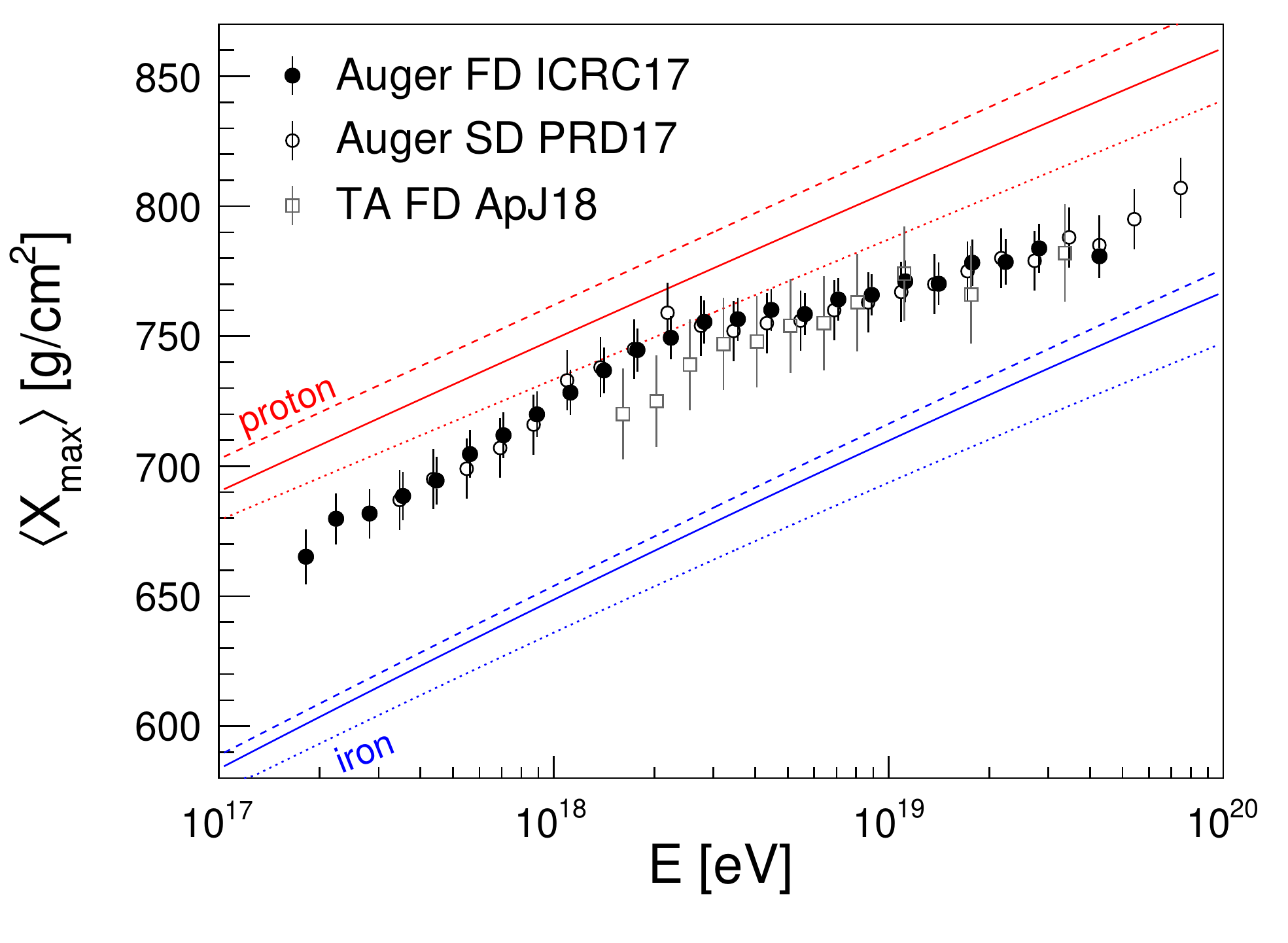}\hspace*{-0.4cm}
  \includegraphics[width=0.52\linewidth]{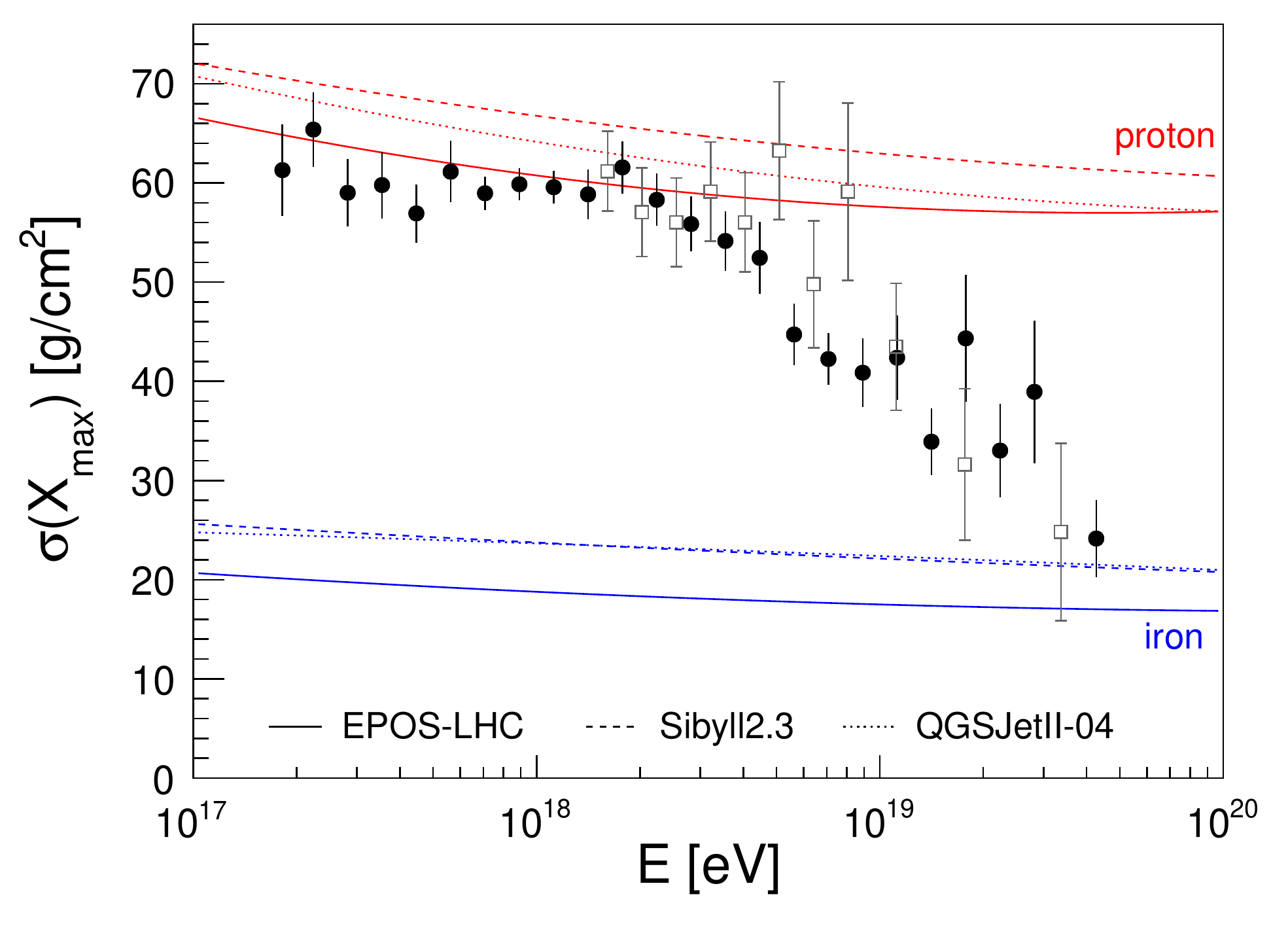}
  \caption{Measurements~\cite{Aab:2017cgk, Bellido:2017cgf, Abbasi:2018nun} of
    the mean ({\it left\/}) and standard
    deviation ({\it right  panel\/}) of the distribution of shower maximum as a
    function of energy.  The
         energy evolution of the mean and standard deviation of \Xmax
         obtained from simulations~\cite{Bergmann:2006yz} are shown as red
         and blue lines; from Ref.~\protect\cite{AlvesBatista:2019tlv}.
         \label{fig:UHECR_composition}    }
\end{figure}

The mass composition of UHECRs can be inferred from  the atmospheric depth
\Xmax where the number of particles in an air shower reaches its maximum.
In the left panel of Fig.~\ref{fig:UHECR_composition},  the \Xmax
values obtained by the PAO using fluorescence detectors (FD) are
presented\footnote{Unfortunately, no update of the common  PAO and TA
  working group on composition was presented at his ICRC.}
as filled dots.
Additionally, results from the surface array (SD) are shown as open dots.
From the evolution of
\Xmax with energy, one can conclude that the composition becomes
lighter between $10^{17.2}$ and $10^{18.33}$\,eV, qualitatively in
agreement with the expectation for a transition from Galactic to
extragalactic CRs in this energy region. Above $10^{18.33}$\,eV,
this trend is reversed and the composition becomes heavier.
The  \Xmax data from  TA shown by squares
are approximately corrected for detector effects by
shifting the mean by +5\,g/cm$^2$~\cite{Yushkov:2018}, as well as
shifted down by 10.4\% in energy. After accounting for these
corrections, the \Xmax data from the two experiments are in good
agreement.
In the right panel of Fig.~\ref{fig:UHECR_composition},  the width \X2
of the \Xmax distributions is shown. Again, the \X2 distribution from
TA has to be corrected for the detector resolution. 
A wide distribution as obtained at low energies can be caused either by
a light or a mixed composition. At higher energies, the distribution becomes
more narrow, pointing to a purer and heavier composition.

\begin{figure}[t]
  \centering
  \includegraphics[width=0.7\columnwidth]{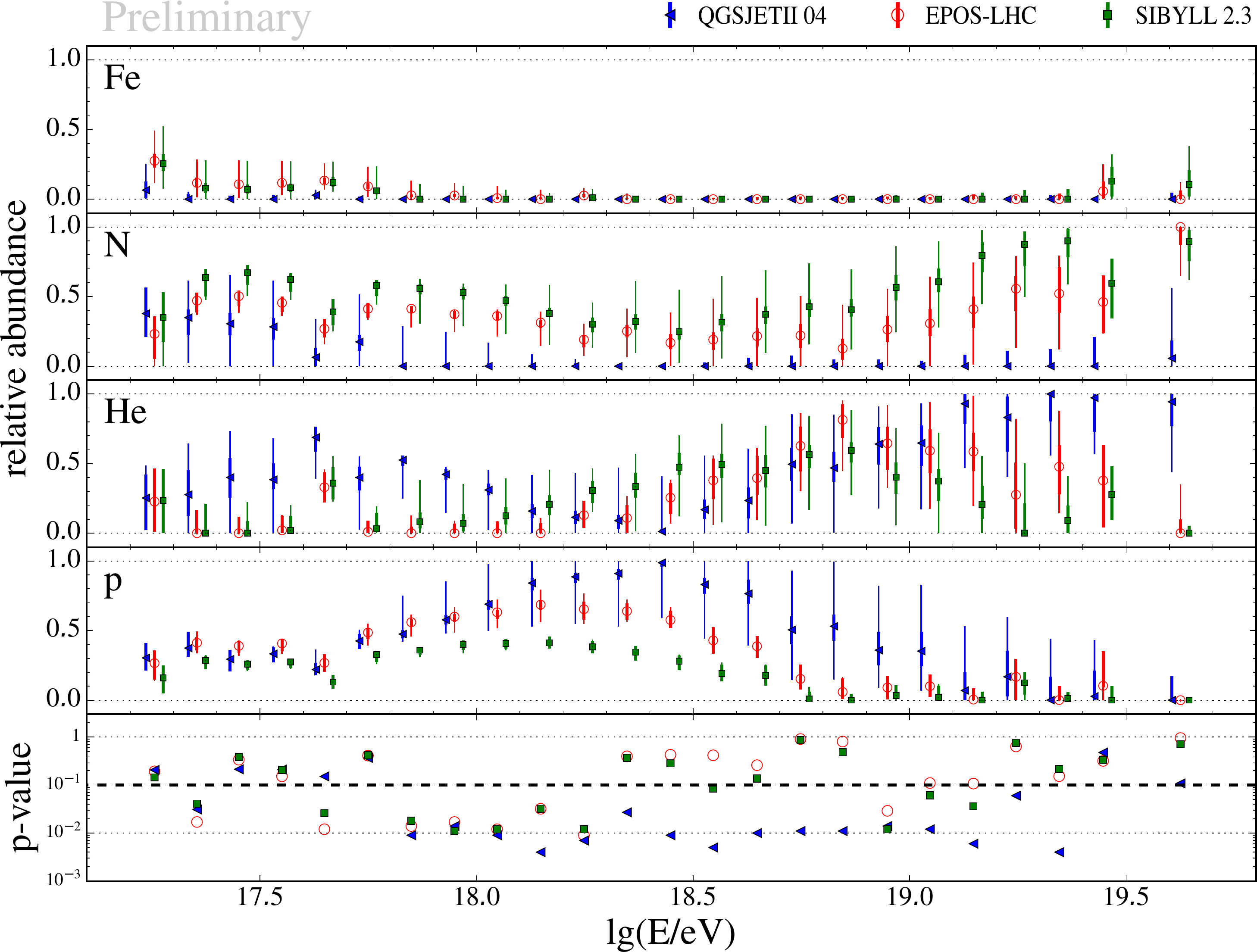}
  \caption{Fraction of the four elements in the UHECR flux;
    from Ref.~\protect\cite{unger:2017cr}. }
\label{fig:auger_composition2}
\end{figure}

Using simulations for hadronic interactions, one can compare
the predicted \Xmax distributions for a mixture of CR nuclei to the
observed  distribution and fit the relative fraction of the  CR nuclei.
The result of such a fit for a mixture of proton, helium, nitrogen and
iron nuclei is shown in Fig.~\ref{fig:auger_composition2}.
Above $10^{18}$\,eV, the dominant component in the UHECR flux changes
successively from protons, to helium and nitrogen, a behaviour
suggestive for the presence of a Peters cycle. At the lowest energies,
there is evidence for a non-zero iron fraction which drops then to zero.

A fit of the combined PAO  data on the energy spectrum and the composition
has been presented in Ref.~\cite{PierreAuger:2020kuy}. In this fit, the
extension of the third segment corresponds to the one expected from
the charge ratio  between CNO amd He. In order to
obtain a clean separation of the mass groups, a very hard
slope\footnote{Steep escape spectra may be generated by threshold effects of
  $A\gamma$ interactions in the source, see, e.g.,
  \cite{Unger:2015laa,Kachelriess:2017tvs}.} of the injection
spectrum into extragalactic space is required: For instance, the fits 
presented in Ref.~\cite{Guido:20218S} use slopes for
$dN/dE\propto E^{-\alpha}$ with $\alpha$ between
$\simeq -1$ and $-2$. Moreover, the
ankle has to be explained by a second extragalactic population which does
not show the same Peters cycle but has a dominantly light composition.
An alternative explanation for the instep is that it is just the most
obvious of several irregularities which are caused by the small number of
sources contributing to the high-energy end of the energy spectrum.

\subsection{Anisotropies and correlations}

\paragraph{Expectations}

A priori, four sub-classes of possible anisotropies were expected to be
imprinted on the arrival directions of extragalactic cosmic rays (see
e.g.\ Ref.~\cite{Kachelriess:2006ip}): 
i) At energies high enough such that deflections in magnetic fields are
sufficiently small, UHECR sources may reveal themselves as small-scale
clusters of UHECR arrival directions.  This requires a  low
density of UHECR sources so that the probability to observe several events
of at least a subset of especially bright sources is large enough. ii) At
lower energies, the energy-loss horizon of UHECRs and thereby the number of
sources visible increases. At the same time, deflections in magnetic fields
become more important. While thus individual sources cannot be identified
anymore,  anisotropies on medium scales should reflect the inhomogeneous
distribution of UHECR sources which in turn is connected---although in a
biased way---to the observed LSS of matter. iii) At even lower energies,
also anisotropies on medium scales disappears, both because the
inhomogeneities in the source distribution will be averaged out because of
the increased energy-loss horizon of UHECRs and because of deflections.
Thus the CR sky appears isotropic, except for a dipole anisotropy
of 0.6\% induced by the cosmological Compton-Getting (CCG)
effect~\cite{Kachelriess:2006aq}. iv) Finally, the Galactic magnetic
field (GMF) can induce anisotropies
in the observed flux of extragalactic UEHCRs (even if it is isotropic at the
boundary of the Milky Way), for rigidities low enough that regions
disconnected from spatial infinity in the CR phase space exist. According to
the estimate of Ref.~\cite{Kachelriess:2005qm}
anisotropies of this kind should be expected in models that invoke a
dominating extragalactic  component already at $E/Z\simeq 4\times 10^{17}$\,V,
if the GMF contains a dipole component.

Obviously,
it is not guaranteed that all these four anisotropies can be observed. If
the average charge of UHECR primaries, their source density and/or
extragalactic magnetic fields (EGMF) are
too large, the integrated flux above the energy where point sources become
visible may be too small for the present generation of UHECR experiments.
Similarly, the presence of a magnetic horizon may replace the CCG effect
by one due to the peculiar flows of nearby galaxies.
It was conjectured that the experimentally easiest
accessible anisotropies are the medium-scale anisotropies connected to the
LSS of UHECR sources~\cite{Cuoco:2005yd,Kachelriess:2005uf}.
However, at present, the only detected anisotropy with more than
$5\,\sigma$ is the dipole moment which is the anisotropy most robust
against deflections. Clearly, the impact of magnetic fields is
more important than expected, partly because the composition
is not proton-like as assumed earlier.

\paragraph{Dipole anisotropy}

The dipole anisotropy $\delta$ (projected into the equatorial plane)
and its phase have been measured starting from TeV energies.
Below $10^{17}$\,eV, the phase of the anisotropy is
approximately constant, except for a flip at $\simeq 200$\,TeV. By contrast,
at energies above $10^{17}$\,eV, the phase changes smoothly towards
100\,degrees in right ascension (R.A.), i.e.\ it points roughly towards
the Galactic anticentre.
This  suggests that the extragalactic CR flux starts at $10^{17}$\,eV
to become sizable. The strength of the dipole at higher energies could
be measured only in 2017 for the first time by the PAO: Performing a
one-dimensional harmonic analysis
in R.A.\ and splitting the events in two energy bins, 4--8\,EeV and
$>8$\,EeV~\cite{Aab:2017tyv,Aab:2018mmi},  the
amplitude $6.5^{+1.3}_{-0.9}\%$ in the second energy $>8$\,EeV deviates
by more than 5\,$\sigma$ from isotropy.
Thus the observed amplitude of the dipole anisotropy is a factor ten
larger than the one expected for the CCG effect.
The estimated direction may be connected to an
overdensity in the local galaxy distribution, seen e.g.\ in the 2MRS
catalogue~\cite{Erdogdu:2005wi}. 
Higher-order harmonics like the quadrupole moment
are consistent with isotropy.
Combining data from PAO and TA allows
one to derive the multipole moments without imposing additional
hypotheses~\cite{TelescopeArray:2021ygq}: The results are compatible
with those using only PAO data, but with uncertainties about twice as
small for the $z$ components of the dipole and quadrupole moment.

\paragraph{Medium scale anisotropies}

\begin{figure}
    \centering
    \includegraphics{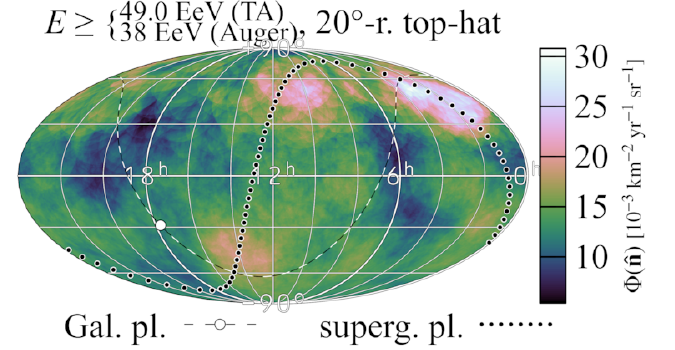}
    \hfil
    \includegraphics{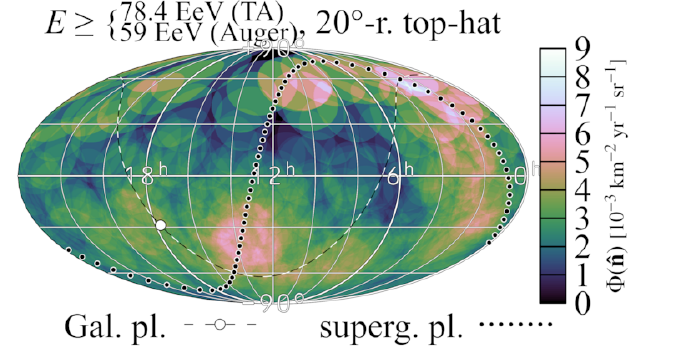}
    \caption{The UHECR intensity in equatorial coordinates averaged over
      $20^\circ$, left above $E=38$\,EeV using the PAO energy scale,
      right above 59\,EeV, from Ref.~\cite{diMatteo:2021b6}. 
    \label{fig:UHECRmap}}
\end{figure}

In Fig.~\ref{fig:UHECRmap}, we show a sky map combining events  above two
selected energy thresholds, averaged over $20^\circ$-radius top-hat windows,
from the TA and PAO experiments~\cite{diMatteo:2021b6}. 
While the significance feature of the the ``TA hotspot'', which is centered
at ${\rm R.A.}\simeq 150^\circ$ and ${\rm dec}\simeq 40^\circ$, has diminished
relative to the ICRC-2017, the two additional ``warm spots''  in the
field-of-view of the PAO have become more prominent.  One of them coincides
with Cen~A and both overlap with the supergalactic plane.

\paragraph{Correlations}

An alternative method to search for UHECR sources are correlation
studies. The recent analysis~\cite{Aab:2018chp} of the PAO data showed
evidence for a correlation of the arrival directions of UHECRs with starburst
galaxies, i.e.\ galaxies which are characterized by exceptionally high
rates of star formation.  Specifically, for UHECR with observed energies
$E>39$\,EeV, a model which attributes 9.7\% of the UHECR flux to nearby
starburst galaxies
(and the remaining 90.3\% to an isotropic background) was found to be favoured,
with $4\sigma$ significance, over the isotropic hypothesis.
About  90\% of the anisotropic flux was found to be associated to four nearby
starburst galaxies: NGC~4945, NGC~253, M83, and NGC~1068.
Alternatively, a correlation analysis with 17~bright nearby AGNs was
performed. Here, the warm spot is related to Cen~A and around 7\% of the
total flux is attributed to the selected AGNs. 
At this conference, the PA and the TA collaborations presented a combined
search for this signal~\cite{TelescopeArray:2021gxg}. The correlation
between the arrival directions of $12\%_{-3\%}^{+5\%}$~of cosmic rays
detected with $E \ge 38~\EeV$ by Auger and with~$E \gtrsim 49~\EeV$ by TA
and the position of nearby starburst galaxies has a $4.2\sigma$ post-trial
significance, when the directions are smeared over
${15.5^\circ}_{-3.2^\circ}^{+5.3^\circ}$\,angular scales.
Thus this combined analysis has a stronger significance than the Auger-only
data but is still short of the discovery level.
In addition, a weaker correlation with the overall galaxy distribution, or
equivalently with the supergalactic plane, was found.

Note  that some starburst galaxies, such as NGC 1068 and NGC 4945, show a
dominant non-thermal contribution from their central regions, indicating
an active black hole. The presence of such AGN-starburst
composites may affect in turn the interpretation of correlation analyses
like the one of Ref.~\cite{Aab:2018chp}. Moreover, the formal significance of
correlation studies has to taken with a grain of salt, as the look-elsewhere
effect due to unpublished unsuccessful searches is impossible
to quantify. New data from  TAx4 and AugerPrime will be therefore
important to test these correlations.

\subsection{Transition energy}

The question at which energy the transition from Galactic to extragalactic
CRs takes place is essential for the understanding of the requirements on
the acceleration mechanisms of Galactic CR sources as well as for the
determination of the nuclear composition and the injection spectrum of
extragalactic sources. In the past, the transition energy has been usually
associated with one of the two evident features of the UHECR spectrum:
The second knee around $E\simeq (1-5)\times10^{17}$\,eV or the
ankle at $E\simeq3\times 10^{18}$\,eV. The latter choice offered a simple
explanation for the sharpness of the ankle as the cross-over between the end
of Galactic flux and the start of the extragalactic component. It allowed
also for an extragalactic injection spectrum  $Q(E)\propto 1/E^\beta$
with $\beta\approx 2$, i.e.\ close to the  expectation for diffusive
shock acceleration. However, this suggestion clearly challenges
acceleration models for Galactic CR sources. Moreover,
this solution leads to the following ``coincidence problem'': Since
the acceleration and diffusion of CRs depends only on rigidity, the end of
the Galactic CR spectrum consists of a sequence of cutoffs at
$ZE_{\max}$, i.e.\ it follows a Peters cycle. If the knee in the total CR
spectrum marks the suppression of Galactic CR protons, then the second knee
should corresponds to the iron knee. Thus, in this interpretation, the
second knee signals the end of the Galactic iron flux from those sources
which contribute the bulk of Galactic CRs.
Therefore an additional Galactic population of CR sources would be required
to fill the gap between the second knee and the ankle. If this population is
unrelated to the standard population of Galactic CR sources, it is surprising
that the normalisation and the slope of the two fluxes is so similar.

Models identifying the second knee as the transition to extragalactic CRs
have to postulate two different extragalactic source populations,
implying again the ``coincidence problem'' described above. This problem
may be avoided if a source type consists of two sub-populations.
Alternatively, these models have to implement
a physical mechanism which explains the ankle
as a consequence of either the propagation of extragalactic CRs or of 
interactions in their sources. The first successful model of this kind,
the dip model~\cite{Berezinsky:2005cq}, requires an almost pure ($\gsim 90\%$)
proton flux and is therefore excluded by composition measurements.
Viable alternative models which rely on interactions of CR nuclei
inside CR sources have been proposed~\cite{Unger:2015laa,Kachelriess:2017tvs}.

How can these two options for the transition energy, the second knee and the
ankle, be experimentally distinguished? It is natural to expect that the
nuclear composition of Galactic and extragalactic CRs should differ, because
of propagation effects and of the different nature of their sources.
In particular, 
the Galactic CR spectrum should become close to its end iron-dominated.
A similar behaviour is expected for the extragalactic flux, shifted however
to higher energies. Thus one expects the extragalactic composition at the
transition energy to be lighter than the Galactic one. Therefore the signature
of the transition in the composition is the disappearance of the (Galactic)
iron, and the increase of a light or intermediate  extragalactic component.
Using only the composition data, the limits on the iron fraction
from Fig.~\ref{fig:auger_composition2} imply that the Galactic contribution
to the observed CR spectrum has to die out before $7\times 10^{17}$\,eV. 
Combining composition and anisotropy measurements strengthens this
conclusion considerably: In Ref.~\cite{Giacinti:2011ww},
it was shown that a light (intermediate) Galactic CR flux leads to a dipole
of order $20\%$ (10\%), overshooting clearly the limits which are on the
percent level~\cite{Aab:2018mmi}. Thus the dominant light-intermediate
contribution to the CR flux measured by the PAO above $3\times 10^{17}$\,eV
has to be extragalactic. Finally, we recall that the smooth change of the
dipole phase at energies above $10^{17}$\,eV towards the Galactic anticentre
supports the suggestion that the transition from Galactic to extragalactic CRs
starts at $10^{17}$\,eV. Thus we conclude that the second knee marks the
transition  between Galactic and extragalactic CRs. Then it is natural
that the second knee is close to its upper end of the range
$(1-5)\times 10^{17}$\,eV of values
considered in the literature, what agrees with the value
$E\simeq (5.0\pm 0.8)\times10^{17}$\,eV determined in
Ref.~\cite{PierreAuger:2020qqz}.

\subsection{Secondary photon and neutrino fluxes}

High-energy cosmic rays can interact with gas or photons in their
sources, and with photons from the extragalactic background light (EBL)
during propagation. The production of neutrinos is by isospin symmetry
tied to the one of photons, and both depend in turn on the flux of primary CRs.
Therefore the observation of these CR secondaries can provide important
information on extragalactic CRs.

Cosmogenic neutrinos are
mostly produced in interactions on EBL photons with energy
$\eps_\gamma\lsim 10$\,eV. With $\langle E_\nu\rangle=E_p/20$, the energy
threshold of this reaction implies that
the flux of cosmogenic neutrinos is suppressed below
$E\approx 2\times 10^{17}$\,eV. If neutrinos are produced by p$\gamma$
interactions in the source, e.g.\
on radiation from an accretion disk with $\eps_\gamma\lsim 1$\,eV, one
expects as threshold $E_{\rm th}\approx 2\times 10^{18}$\,eV.
In contrast, $pp$ interactions lead to a neutrino flux without threshold.
While neutrinos reach the observer unabsorbed, gamma-rays with energies in the
TeV region and above initiate electromagnetic cascades, via the  processes
$\gamma+\gamma_b \to e^++e^-$ and $e^\pm +\gamma_b \to e^\pm + \gamma$~\cite{strong74,Berezinsky:1975zz}.
This cascade develops very fast until it reaches the pair creation threshold.
Thus the Universe acts as a calorimeter for electromagnetic radiation,
accumulating it in the MeV-TeV range as an extragalactic gamma-ray background
(EGRB).
Measurements of the EGRB by Fermi-LAT~\cite{Ackermann:2014usa} were used to
constrain strongly evolving UHECR models~\cite{Kalashev:2007sn,Berezinsky:2010xa,Heinze:2015hhp}.
In particular,  the limit on the allowed energy density of the cascade
radiation, $\omega_{\rm casc}\leq 2 \times 10^{-7}$\,eV/cm$^3$, bounds the
extragalactic neutrino flux~\cite{Berezinsky:2010xa}.

\begin{figure}
  \centering
  \includegraphics[width=0.65\columnwidth,angle=0]{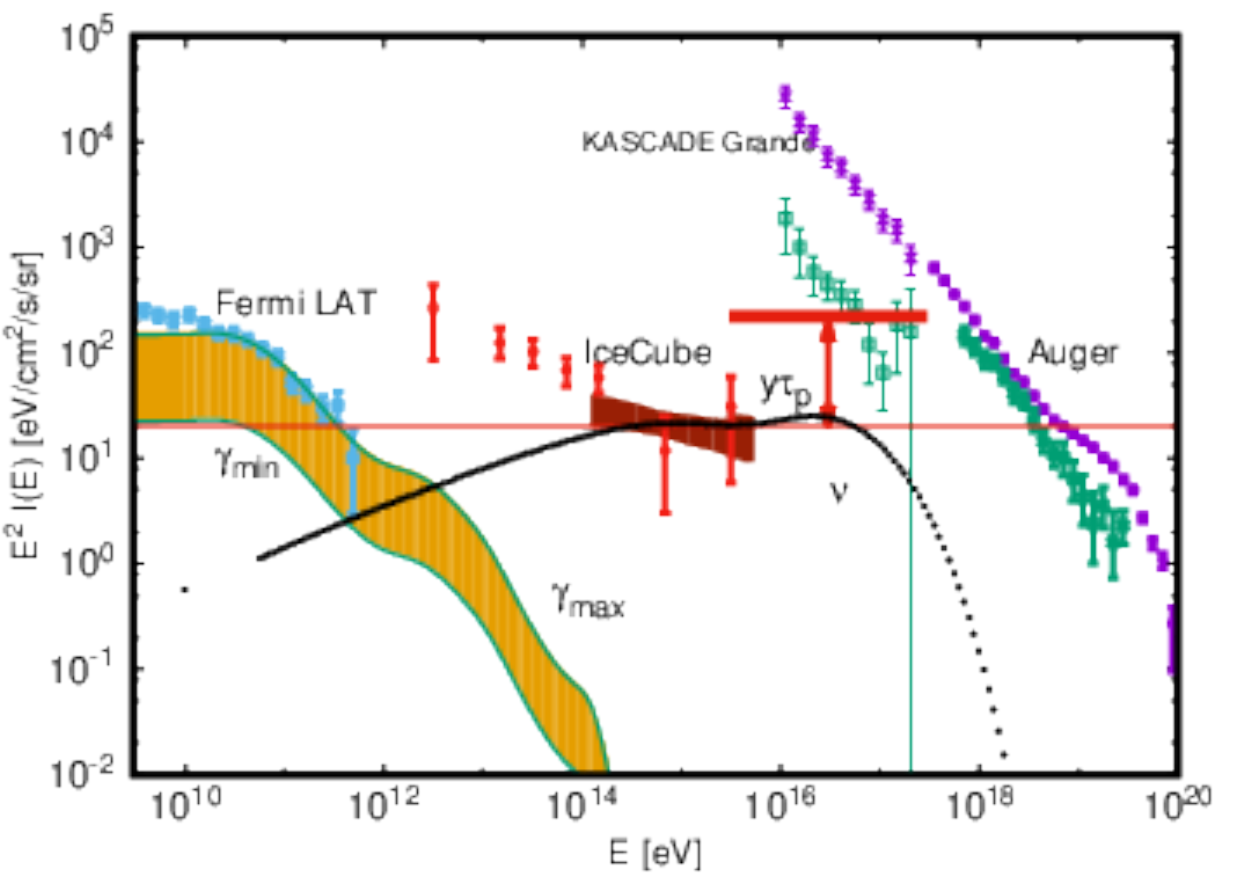}
  \caption{Expected secondary photon and neutrino fluxes from UHECR sources
    compared to the EGRB measured by Fermi-LAT, diffuse neutrino flux by
    IceCube and the all-particle and proton fluxes by KASCADE Grande and
    PAO. The neutrino flux example shown as thin dotted line is from
    Ref.~\protect\cite{Kachelriess:2017tvs}; adapted from
    Ref.~\protect\cite{Kachelriess:2019oqu}. 
\label{fig:secondary}}
\end{figure}

In Ref.~\cite{TheFermi-LAT:2015ykq},
the Fermi collaboration  concluded that up to 86\% of the EGRB is emitted by
unresolved blazars. Taking this result at face value, the room for any
additional injection of photons is very limited. It is therefore desirable
that the same source class explains both UHECRs and the observed neutrino
flux by IceCube. In Fig.~\ref{fig:secondary}, the expectations from
Ref.~\cite{Kachelriess:2019oqu} for the secondary photon and
neutrino fluxes produced by UHECR sources are shown. The EGRB measurements
(blue errorbars)
limit the secondary photon flux from UHECR sources which has a universal
shape indicated by the orange band~\cite{Ginzburg:1990sk,Kachelriess:2011bi}.
The total UHECR flux is shown by violet error-bars, while the green error-bars
give the flux of UHECR protons derived from the PAO and KASCADE-Grande
composition measurements~\cite{unger:2017cr,Apel:2013uni}.
The red points and band represent neutrino measurements by
IceCube~\cite{Aartsen:2016xlq,Aartsen:2017mau}. The horizontal thin red line
shows the average neutrino flux level in the case of
an $1/E^2$ flux, which is of the order of 10\% of the EGRB. In the case of
an $1/E^{2.15}$ neutrino flux, the accompanying photons saturate the gamma-ray
bound.
The thick red line shows the expected level of the proton flux required 
to produce the diffuse neutrino flux. The ratio $y\tau_p$ of the proton
and neutrino fluxes is determined by the corresponding proton interaction
probability  $\tau_p$  and the spectrally weighted average energy
transfer $\langle y\rangle$ to neutrinos. For a spectral slope close
to two,  $y\simeq 0.2$, implying that a large fraction of protons has to
interact inside their sources.
While the high-energy part of the astrophysical neutrino flux is consistent
with $\alpha\simeq 2.1$ and a normalisation close to the cascade bound, at
lower energies a softer
component appears. The sources of this soft component have to be either
extragalactic and hidden, or Galactic but close to isotropically distributed.
Extensive discussions of this connection between extragalactic cosmic
ray sources and the observed high-energy neutrino flux can be found
in the contributions of the Multi-Messenger track at this
conference~\cite{rapp}.

\section{Remarks on the role of magnetic fields}

The strength of the Galactic and extragalactic magnetic fields plays  a
crucial
role for the identification of UHECR sources: In addition to the obvious effect
of deflections, the resulting time-delays determine the effective source
density of transient sources. Moreover, these delays may undermine the
selection of potential UHECR sources based on properties of their
electromagnetic spectrum, if the delays become larger than the activity
periods of the sources.

\paragraph{Galactic magnetic field}

Tess~Jaffe discussed extensively our knowledge about the GMF in her review
talk~\cite{Jaffe}. It is therefore sufficient to
note here that the various GMF models differ by putting emphasis on rotation
measures of extragalactic sources or of Galactic pulsars. The former models
should  therefore provide generally a better description of the magnetic
field in the Galactic halo, while the latter should perform better in the
Galactic plane. However, for none of
these GMF models, the coherence length and the (relative) strength of
the regular and the turbulent field are constrained such that the escape time
of Galactic cosmic rays reproduces the measured secondary-to-primary ratios
below the knee. Consequently, these models cannot be used to investigate, e.g.,
the transition between Galactic and extragalactic cosmic rays or the knee
without adjusting these parameters~\cite{Giacinti:2017dgt}.

\paragraph{Extragalactic magnetic field}

The origin of the EGMF is one of the  outstanding questions in
astrophysics. Its seed fields may be generated in the primordial universe
or by astrophysical processes like galactic plasma outflows. If the field
strength of the EGMF is normalised in both cases
such to reproduce observations in the cores of galaxy cluster, their
filling factors differ drastically. Observationally,
the strength of the EGMF is limited independent of its creation mechanism by
$2\times 10^{-9}$\,G from rotation measures~\cite{Pshirkov:2015tua},
while the present strength of fields with a primordial origin is restricted to
$5\times 10^{-11}$\,G  from CMB anisotropies~\cite{Jedamzik:2018itu}.
The existence of hot spots in the UHECR flux, if they can be firmly
established, provides an alternative way to derive upper limits on the
EGMF. 
There exists also lower limits on the strength and the filling factor of the
EGMF~\cite{Fermi-LAT:2018jdy}. However, it has been argued that they
are invalidated by  plasma instabilities~\cite{Broderick:2018nqf}.

If the average deflection angle per correlation length $L_{\rm c}$ is
small, $R_{\rm L}\gg L_{\rm c}$, the CR propagation resembles a random
walk in the small-angle regime and the variance of the deflection angle
after the distance $d$ is given by
\begin{equation}
\label{eq:thetarms}
 \theta_{\rm rms} \equiv \langle \theta^2\rangle^{1/2}
 \simeq  \frac{(2dL_{\rm c}/9)^{1/2}}{R_{\rm L}} = 25^\circ  Z
 \left( \frac{10^{19}{\rm eV}}{E}\right)
 \left(\frac{d}{100\,{\rm Mpc}}\right)^{1/2}
 \left(\frac{L_{\rm c}}{1\,{\rm Mpc}}\right)^{1/2}
 \left(\frac{B}{10^{-9}{\rm G}}\right) .
\end{equation}
The increased path-length compared to straight-line propagation leads
to the time-delay 
\begin{equation}
\label{eq:Deltat}
 \Delta t \simeq  \frac{d\theta_{\rm rms}^2}{4} = 1.5\times 10^3 \, {\rm yr}\, Z^2
 \left( \frac{10^{20}{\rm eV}}{E}\right)^2
 \left(\frac{d}{10\,{\rm Mpc}}\right)
 \left(\frac{L_{\rm c}}{1\,{\rm Mpc}}\right)
 \left(\frac{B}{10^{-9}{\rm G}}\right)^2 
\end{equation}
of charged CRs relative to photons~\cite{Waxman:1996zn,MiraldaEscude:1996kf}.
This increase  can result in the formation of a magnetic
horizon~\cite{Parizot:2004wh,Berezinsky:2005fa}:
The maximal distance a CR can travel is in the diffusion picture given by
$r_{\rm hor}^2 = \int_0^{t_0} \d t \:D(E(t))$ 
where $t_0$ is the source age. 
If we assume that a magnetic field with correlation length $L_{\rm c}\sim\,$Mpc
and strength $B\sim 
0.1\,$nG exists in a significant fraction of the Universe, then the size
of the magnetic horizon at $E=10^{18}\:$eV is $r_{\rm hor}\sim 100\,$Mpc.
Hence, similar to the GZK effect at high energies, we
see a smaller and smaller fraction of the Universe for lower CR 
energies. As a consequence, the spectrum of extragalactic CRs
visible to us hardens and the extragalactic component becomes suppressed.

\section{Sources, general constraints  and their modelling}

\subsection{Observational constraints}

The observed UHECR intensity fixes the required emissivity ${\cal L}$ of
the UHECR sources,
i.e.\ their energy input per volume and time, up to a model-dependent factor
of order one. For the fit presented in Fig.~\ref{fitspec}, the PAO
determined the emissivity above the ankle,  $E>5\times 10^{18}$\,eV,
at the present epoch as ${\cal L}\simeq 6\times 10^{44}$erg/(Mpc$^3\,$yr). 
If the transition to extragalactic CRs is early, as we argued above,
the corresponding emissivity increases by at least one order of magnitude.
For concreteness, we will use ${\cal L}=1\times 10^{46}$erg/(Mpc$^3\,$yr)
for the following estimates.
In the case of a unique source class, the relation ${\cal L}=Ln_s$
implies that in a  plot of luminosity $L$ vs.\ number density $n_s$ potential
source classes have to sit along a fixed diagonal.
If several source classes contribute significantly to the total emissivity,
they can lie below this diagonal.

The absence of small-scale clustering in the UHECR arrival directions
implies that the source density---or deflections
in magnetic fields---are  sufficiently large. Performing an autocorrelation
analysis of the PAO events with $E>70$\,EeV,
the bounds $n_s>5\times 10^{-4}/$Mpc$^3$ for
a separation angle $\theta=3^\circ$ and  $n_s>6\times 10^{-6}/$Mpc$^3$ for
$\theta=30^\circ$ were derived\footnote{As these limits constitute strong constraints on possible UHECR sources, an update and a more detailed study of the impact of the EGMF are highly desireable.} in Ref.~\cite{PierreAuger:2013waq}.
At these energies, sources should be within 200\,Mpc and one does not expect
from Eq.~(\ref{eq:thetarms}) deflections larger than $20^\circ$  even for the
strongest allowed field. Thus we will use the limit
$n_s>3\times 10^{-5}/$Mpc$^3$
for the density of UHECR sources.
For comparison, the density of the X-ray selected
powerful AGNs with X-ray luminosity $L >10^{43}$erg/s in the energy range
$(0.2-5)$\,keV equals $n_s\sim (1-5)\times 10^{-5}$/Mpc$^3$ within redshift
$z\lsim 0.02$, while the number of  Seyfert galaxies is a factor 20
higher~\cite{Steffen:2003uv}.
Normal galaxies have a density of $n_s\sim 10^{-2}$/Mpc$^3$.

\subsection{Constraints from theory}

Potential sources of UHECRs have to satisfy as two general constraints
the Hillas and Blandford conditions. Combined with measurements of
the present energy density of UHECRs and the absence of small-scale
anisotropies, the Blandford criterium leads to
stringent constraints on the number density of these sources.

\paragraph{Hillas condition}

\begin{figure}
  \centering
  \includegraphics[width=0.49\columnwidth]{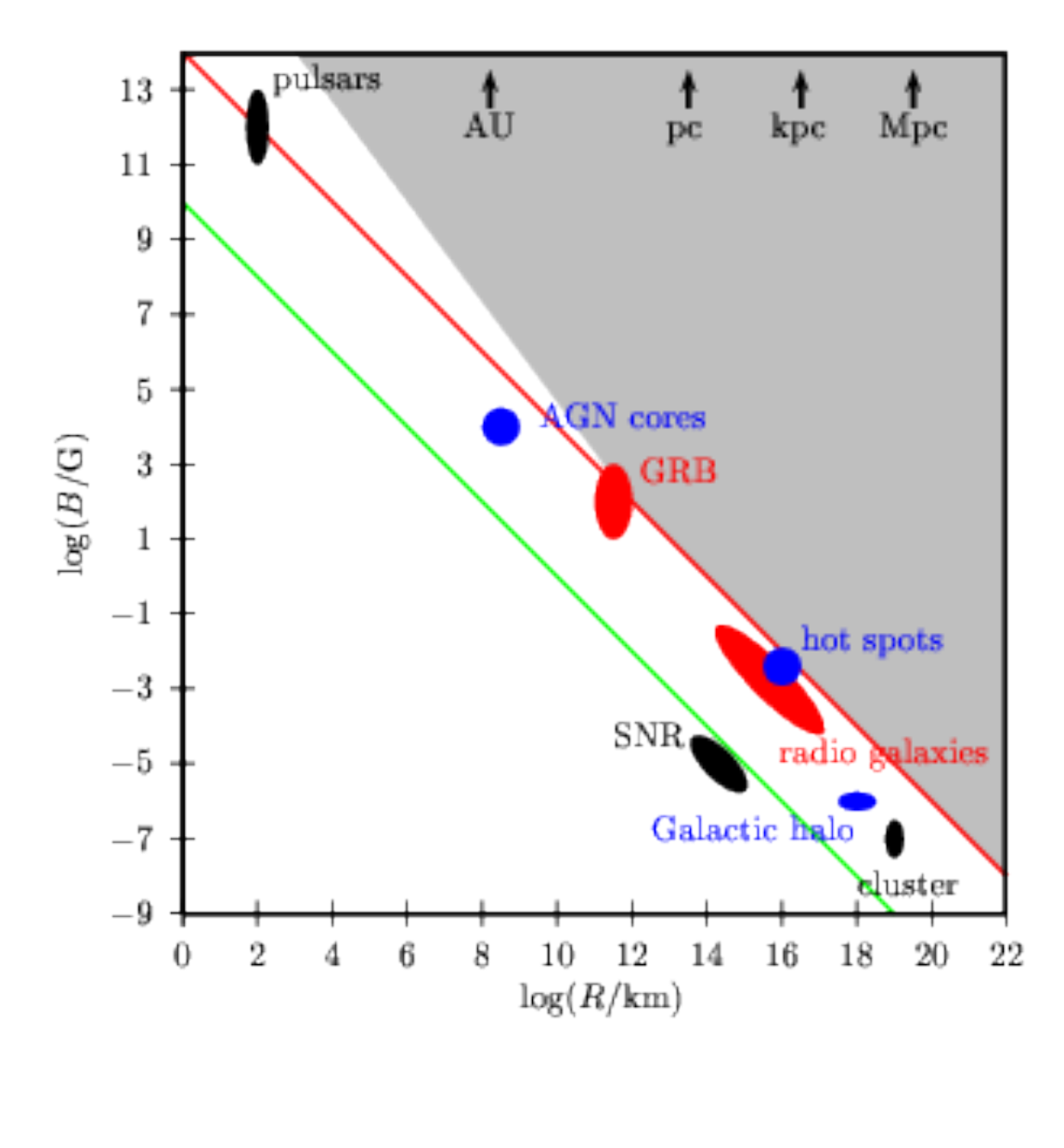}
\caption{\label{Hillas}
Magnetic field strength $B$ versus size $R$ of various  CR sources;
adapted from Refs.~\protect\cite{Kachelriess:2008ze,Ptitsyna:2008zs}.
}
\end{figure}

The Larmor radius $R_L=E/(ZeB)$ of accelerated particles has to fit inside
the accelerator of size $R_s$ which confines the particles by a magnetic
field with strength $B$, i.e.\ $R_L=E/(ZeB)\leq R_s$.  For known magnetic
fields and source sizes, one can constrain thus the maximal achievable
energy as $E_{\max}=\Gamma ZeBR_s$, a constraint shown in the Hillas plot
of Fig.~\ref{Hillas}
for a compilation of potential cosmic ray sources. The Lorentz factor
$\Gamma=(1-\beta^2)^{1/2}$ introduced in $E_{\max}$ accounts for a possible
relativistic bulk motion of the source and is probably only for gamma-ray
bursts and blazars a significant correction.  
Sources able to accelerate protons to $E>10^{20}\,$eV should lie above the 
solid red line, while sources above the green line can accelerate iron up 
to $10^{20}\,$eV. Clearly, a heavy composition of
UHECRs alleviates considerably the acceleration problem of UHECRs.
The main uncertainty in this plot is the strength of the source magnetic
field, since  nonlinear processes typically lead to an amplification of
magnetic fields inside the source. This effect is taken partly into account
in Fig.~\ref{Hillas}, but its exact magnitude is uncertain.
Synchrotron energy losses reduce the allowed area for protons to
the grey area. In addition, the finite acceleration time disfavours large
sources. Thus in general, sources neither too small
(minimizing energy losses) nor too big (avoiding too long acceleration times)
are favored.

\paragraph{Blandford condition}

Constraints on the minimal luminosity of a source able to accelerate
CRs up to a certain energy have been derived for specific source types
and acceleration mechanisms~\cite{1976Natur.262..649L,Blandford:1999hi,Waxman:1995vg,Lemoine:2009vr},
starting from the work of Lovelace~\cite{1976Natur.262..649L} on
radio galaxies.
Blandford stressed first the universality of these limits in
Ref.~\cite{Blandford:1999hi}, and we will call this bound on the luminosity
of a cosmic ray source therefore the Blandford condition.

Let us illustrate this requirement using the simplest example,
the acceleration of charged particles by a regular electric field:
The acceleration of protons to the energy $E=10^{20}\,$eV requires the
potential difference $U=10^{20}\,$V.
What is the minimal power $P$ dissipated by such an accelerator?
In order to use the basic equation $P=UI=U^2/R$ learnt in high-school,
we have to know the appropriate value of the resistance $R$. 
Since an acceleratar operates at densities close to vacuum, we use
$R\sim 1000\:\Omega$ (guided by the ``impedance of the vacuum'',
$R_0=4\pi k_0/c=1/(\epsilon_0 c)\simeq 377\,\Omega$).
Hence a source able to produce protons with $E=10^{20}\,$eV by
acceleration in a regular
electromagnetic field has the minimal luminosity
$L = U^2/R\gsim 10^{37}\,{\rm W} =10^{44}\, {\rm erg/s}$. Including the
effect of possible bulk motions and relaxing the maximal energy, the
bound becomes
\be \label{Bland}
L \gsim 3\times 10^{42}\,{\rm erg/s}  \;\frac{\Gamma^2}{\beta}\,
\left( \frac{E/Z}{5\times 10^{18}\rm eV}\right)^2  .
\ee
How does this derivation apply, e.g., to diffusive shock acceleration?
This case corresponds to a ``circular accelerator'', with an energy gain at
each crossing of the shock front. Microscopically, the electric field 
in the fluid frame which is generated by the plasma flow in the shock region
accelerates charged particles at each crossing of the shock region. Thus
the same argument as above applies.
Note also that $R_0=c\mu_0$ transforms  $P=U^2/R_0$ into an equivalent
(assuming the limit $\beta\sim 1$), often used form  involving the magnetic
energy density.

The bound~(\ref{Bland}) on $L$ can be used to obtain an upper limit on the
density $n_s$ of UHECR sources, since the observed UHECR intensity
fixes the required emissivity ${\cal L}=n_sL$.
Hence, the density of stationary UHECR sources able to accelerate protons to
$E=5\times 10^{18}\,$eV should be smaller than
$n_s< {\cal L}/L\simeq 1\times 10^{-4}/{\rm Mpc}^3 \:\Gamma^{-2}$.

\subsection{Combined constraints on the luminosity and number density of UHECR sources}

\paragraph{Stationary sources}

\begin{figure}
  \includegraphics[width=0.5\columnwidth,angle=0]{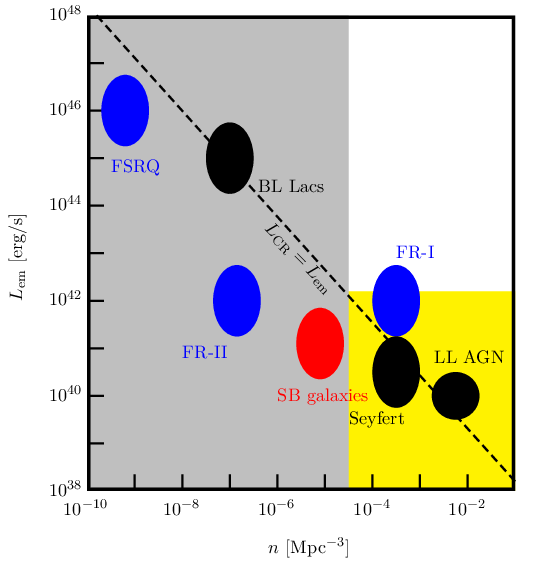}
  \includegraphics[width=0.5\columnwidth,angle=0]{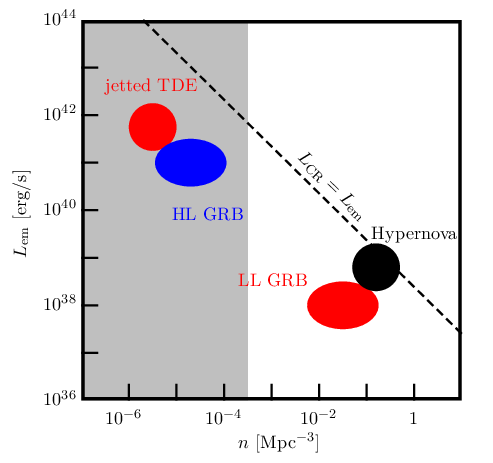}
  \caption{Combined constraints in the luminosity vs.\ number density plot
    shown for stationary (left) and transient (right) sources. The light-gray
    area is excluded by the absence of multiplets, the yellow  area by
    the minimal luminosity argument. \label{fig:blandford}}
\end{figure}

We combine now  in the left panel of Fig.~\ref{fig:blandford} for stationary
sources the lower bound on $n_s$  to avoid small-scale clustering,
$n_s>3\times 10^{-5}/$Mpc$^3$ (for the case of strong EGMFs), the lower
bound on the luminosity and the requirement to reproduce the
observed emissivity.
The first condition excludes the grey area on the left, while the second
requirement excludes (assuming $\Gamma=1$) additionally the yellow area
at the bottom.
Only one source class, FR-I radio galaxies, satisfy both constraints:
Taking into account in addition their non-zero bulk  Lorentz factors,
their luminosity is above the Blandford limit.

How could these conditions be relaxed to allow for more potential
UHECR sources? If a second source populations is responsible
for the the extragalactic CR flux below the ankle, then the requirement
on the emissivity of the first UHECR population is reduced by a factor~20.
In addition, the total energy dissipated in electromagnetic radiation is
larger than in the $X$-ray band used as our proxy, $L_{\rm em}=L_X+\ldots$
Adapting these changes promotes also Seyfert galaxies or low-luminosity
AGNe to possible UHECR sources.

\paragraph{Bursting sources}

The effective number density of bursting sources with rate $R$ and
observed burst duration $\tau$ is given by
$n_s\simeq 3R\tau/5$~\cite{Murase:2008sa}.
The burst duration $\tau$ is dominated by the time delays in magnetic
fields: Using $E=70$\,EeV, i.e.\ the lower energy used to bound $n_s$
from small-scale clustering, together with $B=10^{-11}$\,G gives
$\tau\simeq 6\times 10^{4}$\,yr and
$R=5n_s/(3\tau)\gsim 3\times 10^{-9}$/Mpc$^3$/yr.
The luminosity bound is for bursting sources as GRBs
not a severe restriction, since the true burst time is much smaller than
the observed  duration $\tau$. However, since these sources are rare, they can
typically not provide the observed UHECR emissivity, i.e.\ they
are below the line $L_{\rm CR}=L_{\rm em}$~\cite{Berezinsky:2002nc}.
An exception are hypernovae. In addition, Ref.~\cite{Wang:2007ya} argued that
hypernovae can accelerate protons up to $10^{19}$\,eV, if
their ejecta have a rather flat energy distribution,
$E_k\propto (\beta\Gamma)^{-2}$ (compared to $E_k\propto (\beta\Gamma)^{-5}$
for non-relativistic shocks).
    
\subsection{Comments on the modelling of sources:
  Individual vs.\ average sources}

Combined fits to the energy spectrum and the composition of UHECRs
use often indentical sources with a continuous spatial distribution. In this
idealisation, one replaces sources which typically differ even within
a specific source class in properties like their CR luminosity, maximal
energy, and initial
composition by an ``average source'' plus a redshift evolution of
the emissivity. As an example how this idealisation affects the interpretation
of such fits, one can consider the effect of a
distribution of maximal energies~\cite{Kachelriess:2005xh}: Assuming for
concreteness a power-law distribution for the maximal energies of the
individual sources, $dn/dE_{\max}\propto E_{\max}^{-\beta}$, and a power law
for the energy spectrum of individual sources, $dN_i/dE\propto E^{-\alpha}$,
then the total energy spectrum of all sources becomes
$dN_{\rm tot}/dE\propto E^{1-\alpha-\beta}$ in the limit of
$E_{\max}\to\infty$, if they
are otherwise identical. Thus the inferred slope agrees with
the one of individual sources only for the special case $\beta=1$.

A distribution of luminosities and maximal energies will reduce the
number of active sources at the highest energies. For a smaller number
of sources, the likelihood to find small breaks in the spectrum will
increase. In this picture, it is natural that new features like the
``instep''  show up in the energy spectrum as experimental errors
are reduced.

\section{Specific models}

\subsection{Active galactic nuclei}

\paragraph{Radio galaxies}

Fanaroff-Riley radio galaxies are long-standing candidates for UHECR sources.
In addition to the well-established types of FR-I and FR-II galaxies,
FR-0 galaxies as their low-luminosity extension were suggested in
Ref.~\cite{Merten:2021uvl} as accelerator of UHECRs. These authors argued that
FR-0 galaxies have the required emissivity to power the observed UHECR
energy density, while their rather high density would lead to an almost
isotropic conribution to the UHECR flux. Their  average jet luminosity of
$10^{42}-10^{43}$erg/s is above the Blandford limit. However,
acceleration up to the highest observed energies is possible only
if hybrid acceleration, i.e.\ a  combination of first-order Fermi
with gradual shear acceleration, is realized in these sources.

\paragraph{Blazars}

Gamma-ray blazars have attracted special attention, since IceCube detected
several events from the direction of TXS 0506+056 and PKS 1502+106; see the
highlight talk of Foteini~Oikonomou for an extensive discussion~\cite{FO}.
An example for these works is 
Ref.~\cite{Rodrigues1321}, which modelled the populations of low- and
high-luminosity blazars together with FSRQs. The measured spectrum and
composition of UHECRs  is roughly reproduced in this model above
$10^{18}$\,eV. At the same time, the resulting neutrino fluxes
produced inside the AGN jets can obey the IceCube stacking limits in
the PeV range, while still giving a significant flux of EeV~neutrinos.

\paragraph{NGC~1068}

The Seyfert~II galaxy NGC 1068 has attracted also special attention, because
of a possible excess of IceCube events~\cite{IceCube:2019cia}. At the same
time,
MAGIC~\cite{MAGIC:2019fvw} reported  an upper limit to the gamma-ray flux
above 200\,GeV,
requiring that the gamma-rays accompanying the neutrino flux must be strongly
attenuated in the source.  The authors of Ref.~\cite{Anchordoqui:2021C3} gave
an updated flux prediction for this AGN using the
Stecker-Done-Salamon-Sommers AGN core model
and argued that it can accommodate the IceCube excess.
The authors of Ref.~\cite{Inoue:2021KM} considered instead
the inner regions of the wind launched by the accretion disk.
They found that $pp$ interactions with gas may explain the observations,
if the gas densities are in the range typical for clouds in the
broadline region.
NGC~1068 is not only an AGN, but also a starburst galaxy. The authors of
Ref.~\cite{Eichmann:2021Fn} stressed the need to model both the
AGN core  and the circumnuclear starburst region
in order to provide a complete description of the
non-thermal phenomena of AGN-starburst composite galaxies.

\subsection{Starburst galaxies}

The evidence for a correlation of UHECR arrival directions with the positions
of nearby starburst galaxies poses the question if plausible model for
UHECR acceleration in these galaxies can be developped. A natural acceleration
site is the termination shock of the strong galactic wind expected in
starburst galaxies. The maximal energy  achievable for protons,
$$
E_{\max}\simeq 10^{17}{\rm eV}
\:(t_{\rm acc}/10^9 {\rm yr}) (B/0.3\mu{\rm G}) (v_{\rm sh}/1000\,{\rm km/s}),
$$
depends on the lifetime of the galactic outflows. Typically, they
persist only for times much smaller than the  age of galaxies,
$t_{\rm acc} \sim (10-100)$\,Myr~\cite{Romero:2018mnb},
and thus $E_{\max}$ is too small. The Blandford condition supports this
conclusion. In the model of Ref.~\cite{Peretti:2021oG}, acceleration of
protons to EeV energies require winds with extreme injection power,
$L\sim (10^{44}-10^{45})$erg/s, beyond what is observed in starburst
galaxies.  Consequently, most works on starburst galaxies at this ICRC
concentrated on the acceleration of CRs up to  energies to tens of PeV,
and the associated production of neutrinos, see e.g.\
Refs.~\cite{Condorelli:2021EF,Marinelli:20216t}.

In an alternative view, a starburst galaxy  may not be a
UHECR accelerator as an entity but due to its high rate of GRBs, hypernovae,
or pulsars. An open question is then if starburst galaxies act as
a superposition of single sources, or if the specific environment and
collective effects are
important. In the former case, we recall that the maximal energy achievable
in small sources like pulsars is typically reduced relative to the naive
expectation, while the emissivity of GRBs is too low. Only hypernovae may
satisfy both constraints, and explain in addition the observed
correlations with UHECR arrival directions.

\subsection{Gamma-ray bursts}

Gamma-ray bursts became primary candidates as UHECR and high-energy neutrino
sources, after the so-called $\Gamma^2$ acceleration mechanisms\footnote{A modern incarnation is the espresso mechanism discussed in
Ref.~\cite{Mbarek:2021ST} for AGN jets.}
 was
suggested~\cite{Waxman:1995vg,Vietri:1995hs}. However, the large escape
probability from relativistic shocks after the first cycle restrict the
energy gain per cycle to a factor of order unity, making them much less
effective accelerators as initially
suggested~\cite{Achterberg:2001rx,Lemoine:2006gg}.
Moreover, their emissivity $Q\sim 10^{43}$erg/Mpc$^3$yr is at least a factor
ten too low too explain the observed UHECR flux above the ankle.
Finally, the prediction that GRBs are also sources of high-energy neutrinos
allowed the search of correlations with IceCube events.
From the absence of correlations in the neutrino
data, lower bounds on the number density of the sources can be derived,
excluding effective denisties below $10^{-6}/$Mpc$^3$~\cite{Kowalski:2014zda}.
These constraints from IceCube apply to high-luminosity GRBs and
require either a very low $E_{\max}$ or a small baryon load. Either way,
this excludes high-luminosity GRBs as major UHECR sources.

As a remedy, it was suggested that low-luminosity GRBs, i.e.\ GRBs with
$L\sim (10^{46}-10^{49})$erg/s, may be able to power the UHECR flux. In addition
to the general problem of the too small GRB emissivity, it is unclear
if high enough energies are achievable in these sources which probably
contain shocks in the transrelativistic regime. 
For the special case of GRB~060218, it was argued that effective acceleration
is excluded both in the prompt and afterglow
phase~\cite{Samuelsson:20211n}. On the other hand, the authors of
Ref.~\cite{Rudolph:2021dp} argued for GRB~980425 that heavy nuclei can reach
energies up to $10^{20}$\,eV, if the source is rather extended. This variation
between different low-luminosity GRBs may indicate that they do not form a
single physical source class.

\section{Conclusions}
\label{sec:conclusions}

The field of UHECRs has seen  considerable experimental progress in the
last decade. The energy spectrum is measured precisely and has revealed
a new feature, the instep. The first clear detection of an anisotropy in
the UHECR intensity is a dipole with a $6\%$ amplitude  at energies
$\geq 8\times 10^{18}$\,eV and a direction which points towards the
Galactic anticentre. 
Thus UHECR experiments resolve for the first time the non-uniformity in the
local distribution of UHECR sources. 
Additionally, the PAO claims evidence for a correlation
of the UHECR arrival directions with specific types of UHECR sources,
most significant with starburst galaxies. However, the smearing angle
used in this study is large,
and some confusion between different in source classes (e.g.\ due to
AGN-starburst composites) may occur.

Combining experimental and theoretical constraints on the number density
and luminosity of UHECR sources, luminuous and numerous AGN types as FR-I and
Seyfert galaxies, or alternatively hypernovae, appear as the most promising
UHECR sources. In the latter case, the increased hypernovae rate in starburst
galaxies may explain the suggested correlation of their positions with
UHECR arrival directions.

The determination of the mass composition of UHECRs has progressed too.
The PAO composition data suggest the presence of a Peters cycle
above $2\times 10^{18}$\,eV, with well separated elemental groups
at different energies. Since this scenario requires in the simplest
models very hard injection spectra, it should be scrutinised further.
In particular, after its extension and upgrade, TA should be able 
to confirm or to challenge these results. Moreover, the upgrades of TA and
the PAO are important steps towards the future identification of a proton
(or light) component on an event-by-event basis which is required
to improve correlation analyses with specific source classes.

On the theoretical side, it is desireable to abandon the idea of average
sources and to develop and to employ instead models which include
physically motivated parameters for individual sources. Exploring in
detail the multi-messenger connections
will be another important step towards understanding the sources of UHECRs.
Hypernova explosions and particle acceleration in their trans-relativistic
shocks deserve more thorough studies.

\acknowledgments
\noindent
It is a pleasure to thank all my collaborators, and in particular
Dima Semikoz, for fruitful discussions and work on topics related to this
review. I would like to thank also
Bj\"orn Eichmann and Michael Unger for comments on this article.




\providecommand{\href}[2]{#2}\begingroup\raggedright\endgroup

\end{document}